# The magnetotransport properties of $La_{0.7}Sr_{0.3}MnO_3/BaTiO_3$ superlattices grown by pulsed laser deposition technique

P. Murugavel[a)] and W. Prellier[b)]

*Laboratoire CRISMAT, CNRS UMR 6508, ENSICAEN, 6Bd du Marechal Juin,*

*F-14050 Caen Cedex, France*

**Abstract**

We have investigated the magnetotransport properties of $La_{0.7}Sr_{0.3}MnO_3/BaTiO_3$ superlattices, grown on $SrTiO_3$ substrate by pulsed laser deposition technique, both with current-in-plane and current-perpendicular-to-the-plane directions. Several features indicate the presence of magnetic inhomogeneities at the interfaces which is independent of $BaTiO_3$ layer thickness variation. First, the magnetic property in the superlattices decreases. Second, a hysteresis in magnetoresistance due to the relaxation of the resistive state is observed. Third, a threshold under an applied magnetic field in the magnetoresistance is seen. Such behaviors are in agreement with the phase separation scenario which could be the possible reason for these magnetic inhomogeneities at the interfaces. On the contrary, the magnetoresistance with the current-perpendicular-to-the-plane direction is mostly attributed to the tunneling effect along with the ordering of the spin at the interface. This study confirms the importance of the interfaces in superlattices that can be used to control novel physical properties in oxide materials.

a)Presently at School of physics, Seoul National University, Seoul 151-742, South Korea.
b)Electronic mail: prellier@ensicaen.fr

**I. Introduction**

Rare-earth manganites with hole doping exhibit colossal magnetoresistance properties which make them potential candidates as sensors and memory materials.[1] Particularly $La_{1-x}S_xMnO_3$ is attracting interests because its curie temperature can be adjusted to relatively high values via variation of the Sr concentration. Thin film preparation technique have developed so far, that coherent multilayer with other lattice-matching oxides, in the form of superlattice structure yielded unusual transport properties that cannot be obtained by classical solid-state chemistry route. For example, the multilayer structure formed between half metal like $La_{0.7}Sr_{0.3}MnO_3$ and $SrTiO_3$ reported to have large magnetoresistance in a wide range of temperatures.[2] But the magnetism of the superlattices seems to deviate significantly from the ferromagnetism of the bulk.[3,4] A canted spin structure due to the suppressed ferromagnetic double exchange in competition with the antiferromagnetic super-exchange interactions,[5] the magnetic inhomogeneities due to phase separation near the interface[6-8] and strain induced magnetic surface disorder[9] were being discussed responsible for the observed difference between the superlattice and the bulk. In addition to the magnetoresistance effect, recent reports on superlattices, made of alternating layers of ferromagnetic manganite and ferroelectric perovskite, suggest that there could be a possible magnetoelectric coupling between the two layers in these superlattice structures.[10,11] Therefore we were interested in investigating the interface problem by studying the magnetic and transport properties of superlattices made of ferromagnetic $La_{0.7}Sr_{0.3}MnO_3$

(LSMO) and ferroelectric BaTiO$_3$ (BTO) layers. For this purpose the LSMO/BTO superlattices with varying BTO layer thickness have been made and their magnetotransport properties were measured both in current-in-plane (CIP) and current perpendicular-to-the-plane (CPP) directions and the results were presented in this paper. The results suggest that there exist magnetic inhomogeneities at the interfaces which could be related to the phase separation scenario in the manganite thin films.[6-8]

## II. EXPERIMENTAL

The superlattices of (LSMO$_{10}$/BTO$_3$)$_{25}$, (LSMO$_{10}$/BTO$_6$)$_{25}$, and (LSMO$_{10}$/BTO$_8$)$_{25}$ (hereafter called 10/3, 10/6, and 10/8, respectively) were prepared on polished single crystalline SrTiO$_3$ (100) substrates by a rotating multi-target pulsed-laser deposition technique. For comparison the LSMO film, whose thickness is same as that in superlattices, was also prepared. The films were grown at 720 °C under 150 mTorr of oxygen pressure. After deposition, the sample was cooled to room temperature under 300 Torr of oxygen pressure at the rate of 13 °K/min. 200 mJ of laser power from KrF laser (laser wavelength $\lambda$ = 248 nm) was used for target ablation in all our deposition. The targets were prepared by standard solid state chemistry route using stoichiometric ratios of La$_2$O$_3$, SrCO$_3$, MnO$_2$, BaCO$_3$ and TiO$_2$ as starting materials. The structure of our superlattice samples were analyzed using Siefert 3000P diffractometer (Cu K$_{\alpha1}$, $\lambda$ = 1.5406 Å). The transport properties and the magnetoresistance (MR) of the samples, both in CIP and CPP directions, were measured using a Quantum Design physical property measurement system. Magnetization was measured as a function of temperature and field using a Superconducting Quantum Interference Device magnetometer.

## III. RESULTS AND DISCUSSION

### A. Structural properties

The x-ray $\theta$-$2\theta$ diffraction (XRD) patterns around the (002) fundamental peak of the superlattice films were shown in Fig. 1. The denoted number $i$ indicates the $i$th satellite peak. The presence of numerous satellite peaks, due to the chemical modulation of multilayer structure, proves a formation of well-defined superlattice structure. Superlattice periods, $\Lambda = d_{LSMO} + d_{BTO}$, where d is the respective layer thickness, have been determined from the satellite distances. The laser pulse numbers allowed the estimation of the layer thickness and thus the number of unit cells. We have also carried out the XRD simulation of the superlattice structure using the DIFFAX program[12] and it is found that the experimentally measured peaks are reasonably in good agreement with the simulated one, also shown in Fig. 1. The pseudo-cubic lattice parameter of the BTO (4.033 Å) is larger than that of LSMO (3.876 Å) giving 4 % lattice mismatch. Therefore, as expected the LSMO layers in the superlattice structure could be in highly tensile strained state and it could vary with change in BTO layer thickness.

### B. Magnetic properties

The magnetization of the superlattices and LSMO film measured with respect to temperature from 5 K to 400 K is shown in Fig. 2. The ferromagnetic Curie temperature ($T_C$) of the superlattices (around 240 K) is lower than the parent LSMO film, whose $T_C$ is around 330 K. The reduction of ferromagnetism in manganite thin films and superlattices have been attributed to the spin canting at the interface, phase separation at the interface and strain induced magnetic disorder in the film.[5-9] Though the $T_C$ of our superlattices was lower than the parent LSMO film, they show nearly negligible change with BTO layer thickness. Also the magnetization regarded with various applied magnetic field (shown as an inset in Fig. 2) indicate similar such trend with very small change in saturation magnetization with BTO layer thickness. Although the strain, due to the large lattice mismatch between LSMO and BTO layers, can provide the explanation for the reduced magnetization,[9] the negligible changes in magnetic properties with BTO layer thickness clearly ascertain that the strain variation cannot solely explain the suppressed magnetization shown by the superlattice samples. Recently we reported a near bulk like $T_C$ for the $(LSMO_{10}/BTO_4)_{25}$ superlattice grown under higher oxygen partial pressure compared to the one grown under low oxygen pressure.[13] The results suggest that the suppressed magnetism for the superlattices could be attributed to the presence of magnetic inhomogeneities in samples. In line with the speculation made by Fath et. al., regarding the origins of phase separation in doped manganites;[6] we have indeed provided an evidence of oxygen nonstoichiometry as a possible origin for the magnetic inhomogeneities in the superlattices.[13]

**C. Transport properties in CIP direction**

Fig. 3(a) shows the MR (MR = 100 × $(R_H-R_0)/R_0$, where $R_H$ and $R_0$ is the resistance measured with and without magnetic field, respectively) at 5 K for 10/3, 10/6, 10/8 and LSMO film in CIP direction. The samples were all showing hysteretic MR of significant magnitude. The negative MR continues to increase up to maximum applied field of 7 T, where it shows 22.5, 15, 14 and 7.5 % MR for 10/3, 10/6, 10/8 and LSMO films, respectively. The MR of all superlattices are higher than the parent LSMO film relating an extrinsic contribution to transport at low temperature most likely from the magnetic inhomogeneity[8] and/or spin canting at the interface.[5] Like in ultra thin manganite films where the hysteretic MR with field sweeping has been reported to arise from the time-dependent relaxation of the resistive state due to the presence of magnetic disorder in the film,[14] the hysteretic MR in our superlattice films could be due to magnetic inhomogeneities related relaxation in resistivity.

The inset in the Fig. 3(a) shows resistance versus temperature plot for the 10/3, 10/6, 10/8 and LSMO films measured at zero magnetic fields. The plot shows a clear metal-insulator transition temperature ($T_{MI}$) concomitant with the ferromagnetic $T_C$. Similar to $T_C$, the $T_{MI}$ of all superlattice samples are lower than the LSMO film with almost negligible difference with BTO layer thickness. The coupling of $T_{MI}$ and $T_C$ indicate one-to-one correlation between ferromagnetic and metallic-insulator transition implicit in the double-exchange interaction. The increase in overall resistance with increase in BTO layer thickness seen in Fig. 3(a), confirms the current passing through at least over few LSMO/BTO layers in the superlattice samples in the CIP measurement.

The MR of the samples measured at 100 and 250 K are shown in Fig. 3(b) and 3(c), respectively. The sample 10/3, 10/6, 10/8 and LSMO shows the maximum MR of 25.5, 21, 18, and 9 % and 37, 34, 28, and 24 % at100 and 250 K, respectively. The high MR of 10/3 samples compared to other superlattices could be inferred from the increase in effective number of LSMO/BTO layers along the conduction path due to small BTO layer thickness. The superlattice films were all showing significant value of MR throughout the temperatures well below their $T_{MI}$.

Whereas, the parent LSMO film shows the maximum MR around the $T_{MI}$ (35% at 300 K, not shown in the figure) and below it, the values are negligibly small. We also noticed that there appears a threshold in applied magnetic field above which there is a drastic increase in MR. For example, the inset in Fig. 3(b) shows the threshold field of 1 T in the measured MR of all the samples at 100 K. These results further indicate that there exists a high density of magnetic inhomogeneities at the interface with ordered spins (metallic region) separated by a matrix of disordered region (insulating region).[8] During the field sweeping, initially the field is insufficient to impose large scale spin alignment in the disordered region, but upon reaching a threshold field it appears to allow such ordering to occur with correspondingly large drop in resistance. The microstructural features of the superlattices could clarify the role of the interface quality, chemical mixing, and defects in the transport properties of the superlattices. We believe that the interface between LSMO and BTO would not be clearly visible in the transmission electron microscopy image due to almost equal defocus values and Z values for LSMO and BTO.[15] However, our results suggest that the major contribution for the observed transport properties could come from the presence of the magnetic inhomogeneities due to phase separation scenario at the interface in the LSMO manganite based superlattices, at least in our samples.

**D. Transport properties in CPP direction**

We have also measured the resistance of the samples with respect to temperature and magnetic field in CPP direction. For this purpose the $LaNiO_3$ was used as an electrode prepared in a special geometrical form so as to minimize the geometrical effect arising in junction prepared in usual cross-strip geometry.[16] The resistance versus temperature plot for 10/3, 10/6 and 10/8 samples with 0 T (solid symbols) and 7 T (open symbol) applied magnetic field was shown in Fig. 4(a). The insulating behavior of the superlattice samples down to very low temperature is due to the dominant role of BTO in the resistive contributions from the series combination of metallic LSMO and insulating $BaTiO_3$ along the perpendicular-to-the-plane direction. As expected the resistance of the superlattice increases with increase in BTO thickness. The MR of 10/3 superlattice measured at 5 K is shown as an inset in Fig. 4a (We did not measure the MR for 10/6 and 10/8 samples due to high resistance). The film shows high field MR with hysteretic behavior during field cycling.

At 100 K the 10/3 and 10/6 superlattice shows high field MR of 36 and 13 % as seen in Fig. 4(b). Whereas at 250 K as shown in Fig. 4(c), the 10/3, 10/6 and 10/8 samples shows 12, 9 and 5% high field MR, respectively. Unlike our earlier reports on $Pr_{0.85}Ca_{0.15}MnO_3/Ba_{0.6}Sr_{0.4}TiO_3$ superlattices where we saw an enhancement in MR with increase in $Ba_{0.6}Sr_{0.4}TiO_3$ layer thickness though the parent $Pr_{0.85}Ca_{0.15}MnO_3$ is a robust insulator,[16,17] LSMO/BTO superlattices shows decrease in MR with increase in insulating BTO layer thickness which supports the tunneling effect between the layers. Thus, in CPP geometry, since the conduction path is via same number of interface layer for all the superlattice and the degree of magnetic disorder remains unchanged with BTO layer thickness (inferred from Fig. 2 and the inset in Fig. 3(a)), we believe that the MR of the superlattice and its decreasing trend with increasing BTO layer thickness in CPP measurement could be attributed to the tunneling of spin polarized charge carrier along the thin insulating layers[18] in association with the ordering of the spin in the magnetic inhomogeneous region at the superlattice interfaces under high applied magnetic field.

## IV. CONCLUSION

In conclusion the $(LSMO_{10}/STO_N)_{25}$ superlattice prepared on $SrTiO_3$ substrates exhibits an interesting magnetic and magnetoelectric properties. The suppression of ferromagnetic $T_C$ and metal insulator transition temperature, $T_{MI}$, of the superlattices compared the parent LSMO film indicates there exists strong magnetic inhomogeneities at the interface. Subtle difference in $T_C$ and $T_{MI}$ of the superlattices with respect to the change in BTO layer thickness further elucidate that the degree of magnetic inhomogeneities is independent of BTO layer thickness. The hysteretic MR of the superlattices with field sweeping at low temperature in CIP measurement could be related to time-dependent relaxation of the resistive state due to the magnetic inhomogeneities at the interface. The disorder spins in the inhomogeneous region at the interface imposes a threshold in applied field, above which there is a drastic decrease in resistance due to the ordering of spins resulting in high field magnetoresistance, suggesting a phase separation scenario in the superlattices. The observe MR in CPP direction should be attributed mostly to the tunneling of the charge carriers along with the ordering of the spins at the interface.

## ACKNOWLEDGMENTS

One of the authors (P.M) acknowledges the Ministere de la Jeunesse et de l'Education Nationale for his fellowship (2003/87). We also thank the FAME network of excellence (FP6-500159-1) and the STREP MACOMUFI (033221) and the CNRS for their supports.

**Figure Captions**

Fig. 1. (a) Observed and simulated θ-2θ XRD scan recorded around the (002) reflection of $(LSMO_{10}/BTO_N)_{25}$ superlattices.

Fig. 2. The temperature dependent magnetization of 10/3, 10/6, 10/8 and LSMO films. Magnetization hysteresis loop of 10/3, 10/8 and LSMO films measured at 10 K are shown in the inset.

Fig. 3. (a) The magnetoresistance as a function of field H for 10/3, 10/6, 10/8 and LSMO films measured at 5 K. The inset shows the resistance measured as a function of temperature for the corresponding samples. The magnetoresistance as a function of field H for 10/3, 10/6, 10/8 and LSMO films measured at (b) 100 K and (c) 250 K, in CIP direction. The MR at low field is shown expanded in the inset.

Fig. 4. (a) The resistance measured as a function of temperature for 10/3, 10/6 and 10/8 samples at 0 T (closed symbol) and 7 T (open symbol) field in CPP direction. The inset shows MR as a function of field for 10/3 sample measured at 5 K. The magnetoresistance as a function of field H for the superlattice samples measured at (b) 100 K and (c) 250 K in CPP direction.